\documentclass[aps,preprint,showpacs,superscriptaddress]{revtex4}
\usepackage{amsmath,amssymb,latexsym}
\usepackage[T1]{fontenc}
\usepackage[english]{babel}
\usepackage{graphicx}

\newcommand{\N}{{\mathbb N}}

\begin{document}

\title{Improved bounds in entropic uncertainty relations}
\author{Julio I.\ \surname{de Vicente}}
\affiliation{Departamento de Matem\'aticas, Universidad Carlos III
de Madrid, Avda.\ de la Universidad 30, 28911 Legan\'es, Madrid,
Spain}
\author{Jorge \surname{S\'anchez-Ruiz}}
\affiliation{Departamento de Matem\'aticas, Universidad Carlos III
de Madrid, Avda.\ de la Universidad 30, 28911 Legan\'es, Madrid,
Spain} \affiliation{Instituto Carlos I de F\'{\i}sica Te\'orica y
Computacional, Universidad de Granada, 18071 Granada, Spain}

\pacs{03.65.-w, 03.65.Ta, 03.67.-a}

\begin{abstract}
Entropic uncertainty relations place nontrivial lower bounds to the
sum of Shannon information entropies for noncommuting observables.
Here we obtain a novel lower bound on the entropy sum for general
pairs of observables in finite-dimensional Hilbert space, which
improves on the best bound known to date [Maassen and Uffink, Phys.\
Rev.\ Lett.\ \textbf{60}, 1103 (1988)] for a wide class of
observables. This result follows from another formulation of the
uncertainty principle, the Landau-Pollak inequality, whose
relationship to the Maassen-Uffink entropic uncertainty relation is
discussed.
\end{abstract}

\maketitle

\section{Introduction}

The uncertainty principle states that for quantum systems there is
an irreducible lower bound on the uncertainty in the result of
simultaneous measurements for general pairs of noncommuting
observables. This is one of the key aspects of quantum mechanics,
since it is one of the fundamental points of departure of the theory
with respect to classical physics.

The oldest and most widely used mathematical formulation of the
uncertainty principle is the Heisenberg-Robertson uncertainty
relation \cite{rob}, which places a lower bound on the product of
the standard deviations for any pair of noncommuting observables.
However, two decades ago several authors \cite{deu,hu} pointed out
that this inequality actually fails to express properly the physical
contents of the uncertainty principle, and proposed to use instead
the so-called entropic uncertainty relations (EURs), which place
lower bounds to the sum of the Shannon information entropies of
observables. In fact, for the position-momentum and angle-angular
momentum pairs the optimal (i.e.\ sharpest) EURs were already found
in Ref.\ \cite{bialy}, while in finite-dimensional Hilbert space
several EURs have been derived for general pairs of observables
\cite{deu,kra,maa,opt2}, as well as for particular sets of more than
two observables such as the so-called complementary observables
\cite{sancompl}.

During the last years, EURs in the finite-dimensional setting have
been proved to be not only a subject of fundamental importance, as a
completely rigorous mathematical formulation of the uncertainty
principle, but also a useful tool in quantum information theory. For
instance, EURs have been used to derive separability criteria
\cite{sep}, to show the possibility of locking classical
correlations in quantum states \cite{lock}, and to prove the
security of protocols of quantum cryptography \cite{cryp}.
Unfortunately, the EURs obtained so far for observables in
finite-dimensional Hilbert space are not completely tight in
general, and the optimal lower bound on the entropy sum is only
known in a few special cases. Our aim in this paper is to improve on
the best bound known to date for general pairs of observables acting
on a Hilbert space of arbitrary finite dimension \cite{maa}.

Let $A$ and $B$ denote two Hermitian operators representing physical
observables in an $N$-dimensional Hilbert space, with respective
complete orthonormal sets of eigenvectors $\{|a_i\rangle\}$ and
$\{|b_i\rangle\}$ ($i=1,\ldots,N$), and let $|\psi\rangle$ denote
the normalized state vector describing the quantum (pure) state of
the system. For the sake of simplicity, we assume that both $A$ and
$B$ have non-degenerate spectra, so that there are $N$ possible
outcomes for measurements of each observable and the probabilities
$p_i(A,\psi)$, $p_i(B,\psi)$ ($i=1,\ldots,N$) are given by
\begin{equation}\label{probs}
p_i(A,\psi)=|\langle\psi|a_i\rangle|^2\,,\quad
p_i(B,\psi)=|\langle\psi|b_i\rangle|^2\,.
\end{equation}
An entropic uncertainty relation (EUR) for the pair $A,B$ is an
inequality of the form
\begin{equation}\label{eur}
H_\psi(A)+H_\psi(B)\geq H_{AB}>0\,,
\end{equation}
where $H_\psi(X)$ is the Shannon information entropy corresponding
to the probability distribution $\{p_i(X,\psi)\}$,
\begin{equation}\label{entropy}
H_\psi(X)=-\sum_{i=1}^Np_i(X,\psi)\ln p_i(X,\psi)\,.
\end{equation}
According to Shannon's information theory \cite{sha}, entropy is the
only rigorous quantitative measure of the uncertainty or lack of
information associated to a random variable. The EUR (\ref{eur})
thus sets a nontrivial lower bound, the positive constant $H_{AB}$,
to the joint (information-theoretic) uncertainty about the outcomes
of simultaneous measurements of $A$ and $B$ in any quantum state
\cite{footnote1}.

As first shown by Deutsch \cite{deu}, an inequality of the form
(\ref{eur}) does indeed exist for any pair of observables that do
not share any common eigenstate, as must be expected from a
satisfactory quantitative expression of the uncertainty principle.
Specifically, in Ref.\ \cite{deu} Deutsch proved that
\begin{equation}\label{deutsch}
H(A)+H(B)\geq -2\ln\left(\frac{1+c}{2}\right)\,,
\end{equation}
where
\begin{equation}\label{c}
c = c(A,B) \equiv \max_{i,j}|\langle a_i|b_j\rangle|
\end{equation}
is usually called the overlap of observables $A$ and $B$ (notice
that $1 / \sqrt{N} \leq c \leq 1$ in $N$-dimensional Hilbert space).
The Deutsch EUR (\ref{deutsch}) was later improved by Maassen and
Uffink \cite{maa}, who showed that
\begin{equation}\label{m-u}
H(A)+H(B)\geq -2\ln c \,.
\end{equation}
This inequality is the sharpest EUR known to date for a general pair
of observables in finite-dimensional Hilbert space. In the
particular case when $A$ and $B$ are complementary observables
\cite{sch,kra}, i.e. $|\langle a_i|b_j\rangle| = 1 / \sqrt{N}$ for
all $i,j=1,\ldots,N$, the lower bound $\ln N$ given by (\ref{m-u})
is optimal since it is attained whenever the system is in an
eigenstate of either $A$ or $B$ \cite{kra}. Leaving aside this
special case, however, the Maassen-Uffink EUR is not optimal, in the
sense that the lower bound (\ref{m-u}) is not attained for any
quantum state. The problem of finding the optimal EUR for general
(non-complementary) observables turns out to be very difficult, and
up to now it has only been solved in two-dimensional Hilbert space
\cite{opt2}.

Another alternative mathematical formulation of the uncertainty
principle is provided by the Landau-Pollak uncertainty relation,
which states that
\begin{equation}\label{l-p}
\arccos\sqrt{P_A}+\arccos\sqrt{P_B}\geq\arccos c\,,
\end{equation}
where
\begin{equation}\label{l-p-notation}
P_A \equiv \max_i p_i(A)\,, \quad P_B \equiv \max_j p_j(B)\,.
\end{equation}
This inequality was first considered in the quantum setting by
Uffink \cite{maa,uff}, who adapted the original work of Landau and
Pollak on uncertainty in signal theory \cite{lan}. It satisfies some
of the formal requirements proposed by Deutsch \cite{deu} to
characterize general uncertainty relations, and has been used to
derive separability conditions in the framework of quantum
information theory \cite{lpnuestro}. Remarkably, Eq.\ (\ref{l-p}) is
neither weaker nor stronger than (\ref{m-u}), since one can find
probability distributions allowed by the latter but forbidden by the
former, and vice versa \cite{uff}. However, the Landau-Pollak
inequality does not make use of Shannon's entropy as the measure of
uncertainty, so it cannot be considered as a completely rigorous
expression of the uncertainty principle \cite{footnoteextra}.

As shown by Maassen and Uffink \cite{maa,uff}, the Deutsch EUR
(\ref{deutsch}) can be derived from Eq.\ (\ref{l-p}). In the
following we will prove that use of the Landau-Pollak inequality
(\ref{l-p}) actually enables us to obtain a stronger EUR, which
improves even on the Maassen-Uffink EUR (\ref{m-u}) for pairs of
observables such that
\begin{equation} \label{largeoc}
c(A,B) \geq \frac{1}{\sqrt 2} \simeq 0,707 \,.
\end{equation}
As a by-product, our discussion will clarify the conditions under
which the Landau-Pollak inequality places stronger restrictions on
the probability distributions of $A$ and $B$ than the Maassen-Uffink
uncertainty relation.

\section{Minimization of the entropy sum under the Landau-Pollak constraint}

We proceed by finding the minimum of the entropy sum $H(A)+H(B)$
with the constraint given by Eq.\ (\ref{l-p}). To achieve this goal,
we first consider the minima of the entropy $H(X)$ for probability
distributions which have a fixed value $P$ for their maximum
probability. That is, we seek for the minimum values of the
$N$-variable function $H(X)=\sum_{i=1}^Np_i(X)\ln p_i(X)$ with the
constraints $\sum_{i=1}^Np_i(X)=1$ and $\max_ip_i(X)=P$; notice that
the maximum probability can be repeated $M$ times, with $1\leq M\leq
N$, so the last constraint is in fact a set of $M$ constraints
applying for $i=1,\ldots,M$. The solution to this problem can be
found in Ref.\ \cite{fed}, where it is proved that the minimum
values of $H(X)$ are attained for the probability distributions of
the form
\begin{equation}\label{minimalp}
\{ p_i (X) \}_{\textrm{min}} = \left\{\underbrace{P,\ldots,P}_{M
\textrm{ times}},\;1-MP,\!\!\!\!\underbrace{0,\ldots,0}_{N-M-1
\textrm{ times}}\right\},
\end{equation}
the corresponding values of the entropy being then
\begin{equation}
H_{\textrm{min}}(X)=-MP\ln P-(1-MP)\ln(1-MP)\,,
\end{equation}
for whatever values of $M$ and $P$ such that
\begin{equation}\label{condicionMP}
M\leq\frac{1}{P}<M+1\,.
\end{equation}

The previous result enables us to reduce the problem of finding the
minimum of the entropy sum $H(A)+H(B)$ to a simpler one, namely that
of minimizing the two-variable functional
\begin{align}
& \mathcal{H}(P_A,P_B) \equiv H_{\min}(A)+H_{\min}(B) \nonumber \\
& = \sum_{i=A,B}[-M_iP_i\ln P_i-(1-M_iP_i)\ln(1-M_iP_i)]
\label{funcion}
\end{align}
with the constraints
\begin{equation}\label{consa}
M_i\leq\frac{1}{P_i}<M_i+1\quad(M_i\in\N\,, i=A,B)
\end{equation}
and (\ref{l-p}). For convenience, we will find instead the maximum
of $-\mathcal{H}$, by applying to this functional the
(Karush)-Kuhn-Tucker theory for optimization subject to inequality
constraints \cite{kt}.

Let us first exclude the case in which $P_i=1/M_i$ for at least one
$i$, which will be treated separately. The Lagrangian for this
problem is
\begin{align}
\mathcal{L} = & \sum_{i=A,B}\left[M_iP_i\ln
P_i+(1-M_iP_i)\ln(1-M_iP_i)\right.\nonumber\\
& \left. + \mu_i\left(\frac{1}{M_i}-P_i\right)+
\nu_i\left(P_i-\frac{1}{M_i+1}\right)\right]\nonumber\\
& + \lambda(\arccos\sqrt{P_A}+\arccos\sqrt{P_B}-\arccos
c)\,,\label{lagrangian}
\end{align}
where $\lambda,\mu_i,\nu_i\geq0$ are Lagrange undetermined
multipliers. The Kuhn-Tucker necessary conditions \cite{kt} for a
point to be a maximum are then, with $i=A,B$,
\begin{gather}
M_i\ln\frac{P_i}{1-M_iP_i}-\frac{\lambda}{2\sqrt{P_i(1-P_i)}}-\mu_i+\nu_i=0\,,
\label{kt1} \\
\mu_i\left(\frac{1}{M_i}-P_i\right)=0\,, \label{kt2} \\
\nu_i\left(P_i-\frac{1}{M_i+1}\right)=0\,,\label{kt3} \\
\lambda(\arccos\sqrt{P_A}+\arccos\sqrt{P_B}-\arccos c)=0\,,
\label{kt4}
\end{gather}
provided that Eqs.\ (\ref{l-p}) and (\ref{consa}) are still
fulfilled. Notice that the Kuhn-Tucker conditions are necessary for
a point to be a maximum, but they are not sufficient. Therefore,
once we find all the solutions of Eqs.\ (\ref{l-p}), (\ref{consa})
and (\ref{kt1})-(\ref{kt4}) we will have to check which one
corresponds to the actual maximum.

Since we are restricting ourselves to the case when $P_i\neq1/M_i$
for $i=A,B$, we must set $\mu_i=\nu_i=0$ also for $i=A,B$ if we want
Eqs.\ (\ref{kt2}) and (\ref{kt3}) to be compatible with
(\ref{consa}). If $\lambda=0$ as well, then condition (\ref{kt1})
reduces to $P_i=1/(1+M_i)$, which contradicts Eq.\ (\ref{consa}).
Therefore $\lambda\neq0$, so that Eq.\ (\ref{kt4}) reduces to
\begin{equation}\label{l-peq}
\arccos\sqrt{P_A}+\arccos\sqrt{P_B}=\arccos c\,.
\end{equation}
This means that, as it was reasonable to expect, the optimal
probability distribution saturates the Landau-Pollak uncertainty
relation. Using the trigonometric identity
\begin{equation}
\arccos x+\arccos y=\arccos\left(xy-\sqrt{(1-x^2)(1-y^2)}\right)\,,
\end{equation}
Eq.\ (\ref{l-peq}) implies that
\begin{align} \label{condonc}
c&=\sqrt{P_AP_B}-\sqrt{(1-P_A)(1-P_B)}\nonumber\\
&\leq\frac{1-\sqrt{(M_A-1)(M_B-1)}}{\sqrt{M_AM_B}}\,,
\end{align}
where the inequality in the right-hand side follows from
(\ref{consa}). Since $c\in(0,1]$, we see from (\ref{condonc}) that
\begin{equation}\label{arccos}
\min(M_A,M_B)=1\,,\quad c\leq\frac{1}{\sqrt{\max(M_A,M_B)}}\,.
\end{equation}
On the other hand, Eq.\ (\ref{kt1}) yields
\begin{align} \label{lambdaeq}
\lambda&=2M_A\sqrt{P_A(1-P_A)}\ln\frac{P_A}{1-M_AP_A}\nonumber\\
&=2M_B\sqrt{P_B(1-P_B)}\ln\frac{P_B}{1-M_BP_B}\,.
\end{align}

Equation (\ref{lambdaeq}) has several solutions, each of which
provides a possible minimum for $\mathcal{H}$. For instance, if we
assume that $P_A=P_B$, then Eq.\ (\ref{l-peq}) implies that
\begin{equation}\label{igualp}
P_A=P_B=\frac{1+c}{2}\,,
\end{equation}
while (\ref{consa}) and the first equation in (\ref{arccos}) impose
that $M_A=M_B=1$. Therefore, this solution gives the following
candidate for the minimum of $\mathcal{H}$,
\begin{equation}\label{ana}
\mathcal{F}(c) \equiv
-(1+c)\ln\frac{1+c}{2}-(1-c)\ln\frac{1-c}{2}\,.
\end{equation}

If we now assume that $P_A\neq P_B$, we have $M_A=1$, $M_B=M\in\N$
because of Eq.\ (\ref{arccos}). Then the possible minima of
$\mathcal{H}$ come from the solutions of the equation
\begin{equation}\label{key}
\sqrt{P_A(1-P_A)}\ln\frac{P_A}{1-P_A}=M\sqrt{P_B(1-P_B)}\ln\frac{P_B}{1-MP_B}
\end{equation}
for $M=1,2,\ldots$, each of which will be only valid within the
range $c\leq1/\sqrt{M}$ due to the second equation in
(\ref{arccos}). Unfortunately, Eq.\ (\ref{key}) with $P_A\neq P_B$
cannot be solved by analytical means and its solutions must be
calculated numerically. If, recalling (\ref{l-peq}), we define
\begin{equation} \label{deftrig}
P_A\equiv\cos^2\alpha\,,\quad P_B\equiv\cos^2(\theta-\alpha)\,,\quad
c\equiv\cos\theta\,,
\end{equation}
then Eq.\ (\ref{key}) is rewritten as
\begin{align} \label{num}
&\sin2\alpha\ln\left(\frac{1+\cos2\alpha}{1-\cos2\alpha}\right) +
M\sin2(\alpha-\theta) \nonumber \\ & \times \ln
\left(\frac{1+\cos2(\alpha-\theta)}{2(1-M\cos^2(\alpha-\theta))}\right)
= 0\,,
\end{align}
where $\alpha\neq\theta/2,\theta/2+\pi/4$ in order to specify
$P_A\neq P_B$. We will denote by $\mathcal{H}_M(c)$ the possible
minimum of $\mathcal{H}$ obtained by substituting into Eq.\
(\ref{funcion}) the numerical values of $P_i(c)$ corresponding by
means of (\ref{deftrig}) to the solution $\alpha(\theta)$ of Eq.\
(\ref{num}).

Finally, we consider what happens if we allow $P_i=1/M_i$ for $i=A$
and/or $i=B$. Then we get the solution $P_A=1$, $P_B=c^2$, which
yields the possible minimum
\begin{equation}\label{ana2}
\mathcal{G}(c) \equiv -c^2[1/c^2]\ln
c^2-(1-c^2[1/c^2])\ln(1-c^2[1/c^2])\,,
\end{equation}
where $[x]$ denotes the integer part of $x$. There also exist other
solutions which turn out to be uninteresting \cite{footnote2}.

We now have to select between all the previous solutions the actual
minimum of $\mathcal{H}$, which will be our novel lower bound for
$H(A)+H(B)$. For $c\geq1/\sqrt{2}$ we just have three possibilities,
namely the analytical bounds $\mathcal{F}(c)$, $\mathcal{G}(c)$ and
the numerical bound $\mathcal{H}_1(c)$. These three possible bounds
are plotted in Fig.\ 1 together with the Maassen-Uffink bound
(\ref{m-u}). From there we readily see that the actual lower bound
on the entropy sum equals $\mathcal{F}(c)$ when $c\geq
c^*\simeq0.834$, and $\mathcal{H}_1(c)$ when $1/\sqrt{2}\leq c\leq
c^*$. We also see that, in both cases, our lower bound is stronger
than the Maassen-Uffink one. It is worth noting that
$\mathcal{G}(c)$ is not the actual minimum for any value of $c$,
although it is very close to the numerical bound $\mathcal{H}_1(c)$
and practically overlaps with it within a considerable range.

On the other hand, in the case when $c\leq1/\sqrt{2}$ the minimum of
$\mathcal{H}$ fails to improve on the Maassen-Uffink bound, as can
be readily seen from the graph of $\mathcal{H}_1(c)$ displayed in
Fig.\ 1. Nevertheless, since $\mathcal{G}(c)$ interpolates the
Maassen-Uffink bound in the only points in which the latter is
optimal, i.e.\ $c=1/\sqrt{n}$ with $n\in\mathbb{N}$ (see Fig.\ 1),
one could think that the former could indeed be the actual lower
bound on $H(A)+H(B)$. However, it is possible to find examples with
a slightly lower entropy sum that disprove this conjecture.

\begin{figure}[t]
\includegraphics[scale=0.5]{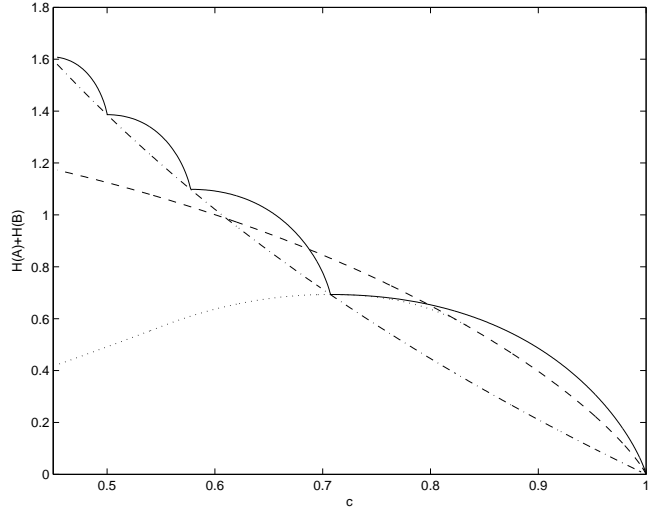}
\caption{Actual and possible lower bounds in the EUR (\ref{eur}) for
a pair of observables with overlap $c$: Maassen-Uffink bound
(dash-dotted line), $\mathcal{F}(c)$ (dashed line),
$\mathcal{H}_1(c)$ (dotted line), and $\mathcal{G}(c)$ (solid
line).}
\end{figure}

\section{Conclusions}

In summary, using the Landau-Pollak inequality (\ref{l-p}) we have
managed to improve the Maassen-Uffink bound for EURs in a
finite-dimensional Hilbert space for the set of observable pairs
that fulfill the large overlap condition (\ref{largeoc}). The
strongest lower bound $H_{AB}$ that is now available for the EUR
(\ref{eur}) corresponding to a general pair of observables with
overlap $c$ can thus be written as the piecewise function
\begin{equation} \label{newbound}
H_{AB} = \left\{ \begin{array}{ccl} -2\ln c & \textrm{if} & 0 < c
\leq 1/\sqrt{2}\,, \\ \mathcal{H}_1(c) & \textrm{if} & 1/\sqrt{2}
\leq c \leq c^*\,, \\ \mathcal{F}(c) & \textrm{if} & c^* \leq c \leq
1\,,
\end{array} \right.
\end{equation}
where $c^*\simeq0.834$, the analytical bound $\mathcal{F}(c)$ is
given by Eq.\ (\ref{ana}), and the numerical bound
$\mathcal{H}_1(c)$ was defined after Eq.\ (\ref{num}).

It is interesting to note that for observables acting on a
two-dimensional Hilbert space, where the large overlap condition
(\ref{largeoc}) always holds, the bound in Eq.\ (\ref{newbound})
coincides with the optimal bound obtained in \cite{opt2}. This fact
implies that in the general (higher-dimensional) case the bound in
(\ref{newbound}) is the best possible bound that can be expressed in
terms only of the overlap $c$. As a by-product, our derivation shows
that the Landau-Pollak inequality is optimal in the two-dimensional
case.

Our derivation also shows that the Landau-Pollak uncertainty
relation is stronger than the Maassen-Uffink EUR for observables
that fulfill the large overlap condition (\ref{largeoc}). It is
remarkable that an inequality based on such a simple measure of
uncertainty, which ignores all but one of the values of the
probability distribution, exhibits this strength. On the other hand,
for observables that do not satisfy condition (\ref{largeoc}), the
Landau-Pollak inequality turns out to be weaker than the
Maassen-Uffink EUR. As a matter for future research, it would be
interesting to check whether other formulations of the uncertainty
principle relying on different measures of uncertainty (see Ref.\
\cite{uff}) can be used in an analogous way to derive better bounds
for EURs.

\section*{Acknowledgments}

We acknowledge support by Direcci\'on General de Investigaci\'on
(Ministerio de Educaci\'on y Ciencia) under grant No.\
MTM2006-13000-C03-02 and by Universidad Carlos III de Madrid and
Comunidad Aut\'onoma de Madrid (project No.\ CCG06-UC3M/ESP-0690).
J. S.-R. was also supported by the DGI (MEC) grant FIS2005-00973,
and the Junta de Andaluc\'ia research group FQM-0207.


\begin{thebibliography}{widest-label}
\bibitem{footnote1} Notice that the concavity of the entropy
functional implies that any inequality of the form (\ref{eur}),
while stated only for pure states, remains also valid for general
(mixed) states. In what follows, for the sake of brevity, we do not
write explicitly the dependence of probability distributions and
entropies upon the quantum state of the system.
\bibitem{footnoteextra} In particular, contrary to what happens for
EURs (cf.\ \cite{footnote1}), the uncertainty functional in Eq.\
(\ref{l-p}) is not concave, so that the validity of this inequality
does not extend in an obvious way from pure to general (mixed)
states.
\bibitem{footnote2} If only $P_A=1/M_A$, then $\mu_B=\nu_B=0$ and Eqs.\
(\ref{l-peq})-(\ref{arccos}) hold like in the previous cases. The
analogue of Eq.\ (\ref{kt1}) for $i=A$ now imposes that $M_A=1,2$ in
order to have non-negativity of the Lagrange multipliers. The
solution for $M_A=1$ gives Eq.\ (\ref{ana2}), which is thus proven
to be valid in principle for all values of $c$, while the solution
for $M_A=2$ can only hold when $c\leq1/\sqrt{2}$. If $P_i=1/M_i$ for
both $A$ and $B$, a similar reasoning yields $M_i=1,2$ for $i=A,B$,
and the resulting bounds are clearly non-optimal.
\end{thebibliography}
\end{document}